\documentclass[graybox]{svmult}

\usepackage{mathptmx}       
\usepackage{helvet}         
\usepackage{courier}        
\usepackage{type1cm}        
\usepackage{makeidx}         
\usepackage{graphicx}        
\usepackage{multicol}        
\usepackage[bottom]{footmisc}

\usepackage{amsmath}

\definecolor{mygreen}{rgb}{0,0.5,0} 
\definecolor{myorange}{rgb}{1.,0.5,0} 
\definecolor{myblue}{rgb}{0,0,0.75} 
\definecolor{mymagenta}{cmyk}{0,1,0,0.12}

\newcommand{\gtext}[1]{{\color{mygreen}#1}}

\usepackage{color}
\definecolor{keywordcolor}{rgb}{0,0,0} 
\providecommand{\keyword}[1]{\index{#1}{\color{keywordcolor}#1}}

\makeindex             

\begin{document}

\title*{Engineering of quantum dot photon sources via electro-elastic fields}
\author{Rinaldo Trotta and Armando Rastelli}
\institute{Institute of Semiconductor and Solid State Physics \at Johannes Kepler University Linz, \at Altenbergerstrasse 69, Linz, A-4040, Austria 
\and
 \email{rinaldo.trotta@jku.at}}
%
%
\maketitle


\abstract{The possibility to generate and manipulate non-classical light using the tools of mature semiconductor technology carries great promise for the implementation of quantum communication science. This is indeed one of the main driving forces behind ongoing research on the study of semiconductor quantum dots. Often referred to as "artificial atoms", quantum dots can generate single and entangled photons on demand and, unlike their natural counterpart, can be easily integrated into well-established optoelectronic devices. However, the inherent random nature of the quantum dot growth processes results in a lack of control of their emission properties. This represents a major roadblock towards the exploitation of these quantum emitters in the foreseen applications.\newline
This chapter describes a novel class of quantum dot devices that uses the combined action of strain and electric fields to reshape the emission properties of single quantum dots. The resulting electro-elastic fields allow for control of emission and binding energies, charge states, and energy level splittings and are suitable to correct for the quantum dot structural asymmetries that usually prevent these semiconductor nanostructures from emitting polarization-entangled photons. Key experiments in this field are presented and future directions are discussed.
\newline\indent
}

\section{Engineering of quantum dot photon sources via electro-elastic fields}
\label{sec:1}

Initially referred to as "quantum boxes" \cite{Cibert1986}, semiconductor quantum dots (QDs) are nanostructures made of several thousands of atoms that can self-assemble during hetero-epitaxial growth \cite{michler2003single}. QDs are capable to confine the motion of charge carriers in three dimensions and feature discrete energy levels. The latter property, which is a direct consequence of the laws of quantum mechanics, has earned QDs the well-known nickname of "artificial atoms". Looking at these nanostructures with the eyes of a passionate material scientist, there is no doubt that QDs represent one of the most spectacular examples of our ability to manipulate matter at the atomic scale, result of more than 50 years of extensive research in semiconductor and solid state physics. In the last 15 years, however, the interest for QDs has pushed its boundaries into the realm of quantum optics. Seminal works demonstrated that QDs can act as triggered sources of single photons \cite{Michler22122000} and entangled photon pairs \cite{PhysRevLett.84.2513,PhysRevLett.96.130501,stevenson2006semiconductor} and can be easily integrated into conventional optoelectronic devices~\cite{salter2010entangled} and optical microcavities~\cite{Reithmaier2004,Hennessy2007,dousse2010ultrabright}. The appealing idea of exploiting semiconductor-based sources of non-classical light for quantum technologies has thereby triggered efforts of researchers working at the interface between quantum optics and condensed-matter physics. Nowadays, the "quality" of the single and entangled photons produced by these nanostructures is reaching levels comparable to trapped atoms \cite{he2013demand,muller2014demand} or parametric down-converters \cite{trotta2014highly}, and advanced quantum optics experiments, such as quantum teleportation \cite{nilsson2013quantum,gao2013quantum}, have been recently performed.

In spite of these accomplishments, the establishment of QD photon sources as viable building blocks for quantum communication requires a number of extraordinary challenges to be overcome. The need of Fourier-limited \cite{he2013demand,muller2014demand}, bright \cite{dousse2010ultrabright,claudon2010highly,versteegh2014observation}, and site-controlled \cite{juska2013towards,trotta2011fabrication} photon sources remains certainly a problem, and some of the groundbreaking results that have been recently achieved are discussed within this book. There is another issue, however, which becomes crucial as soon as the number of quantum emitters required for the envisioned application increases: different from real atoms, each QD possesses its size, shape, composition \cite{rastelli2008three} and, as consequence, a unique emission spectrum. This hurdle is a direct consequence of the stochastic growth processes and has a dramatic effect, e.\ g.\ on the capability of transferring quantum information between distant QD-based qubits \cite{Gisin2002Quantum}. To better explain this point, let us consider Hong-Ou-Mandel two-photon interference \cite{hong1987measurement} between photons emitted by two remote QDs \cite{patel2010two,Flagg2010interference}, a key operation of existing protocols of large-distance quantum communication \cite{duan2001long}. This quantum-mechanical phenomenon consisting in the coalescence of two single photons into a two-photon collective state can be observed when single photons impinge onto a beamsplitter. There, the photon wavepackets should be indistinguishable in all the possible degrees of freedom. While polarization and space overlap can be easily achieved, if we restrict our discussion to Fourier-limited QD photons (see chapter by A. Kuhn) the overlap in energy is what eventually reduces the visibility of two-photon interference. Considering that the inhomogeneous broadening of QD emission is typically tens of meV, the probability of finding two QDs for which photons have the same energy within typical radiative-limited emission linewidth ($ \sim \mu$eV) is $< 10^{-4}$. The situation worsens when one requires entangled photons to impinge onto \gtext{a} beamsplitters, i.e., when one aims at quantum teleportation or entanglement swapping between distant nodes \cite{pan1998experimental}. In fact, the capability of QDs to generate photon pairs that have high enough level of entanglement to violate Bell's inequality \cite{ekert1991quantum} is hindered by the presence of structural asymmetries, which manifest themselves, via the anisotropic electron-hole exchange interaction \cite{bayer2002fine}, in the appearance of an energetic difference between the two bright excitonic states, the well-known fine structure splitting (FSS). When the FSS is larger than the radiative-limited emission linewidth of the excitonic transition ($ \sim 1~\mu$eV), the entanglement is strongly reduced \cite{hudson2007coherence}. Recent theoretical calculations \cite{gong2014statistical} show that only a very low portion of as-grown QDs are free of asymmetries (1 over 1000 for standard Stranski-Krastanow QDs) and the numbers increase slightly if very sophisticated growth protocols are employed \cite{juska2013towards}. Therefore, the probability of finding two as-grown QDs suitable to swap entanglement is considerably small ($10^{-9}$ or less), and it is practically zero in the case of experiments involving several sources. Therefore, the future of QDs for applications critically depends on our capability to precisely control their optical properties within tolerances which are too small to be met even by the most refined fabrication methods. These hurdles have naturally led to the idea of post-growth tuning of the QD emission via the application of external perturbations, such as electric fields \cite{bennett2010electric,gerardot2007manipulating,vogel2007influence,krenner2005direct}, elastic stress \cite{seidl2006effect,kuklewicz2012electro,zander2009epitaxial}, magnetic \cite{bayer2002fine,hudson2007coherence}, and optical \cite{muller2009creating} fields. Among the others, piezoelectric-induced strains \cite{zander2009epitaxial} and vertical electric fields \cite{bennett2010electric} are among the most promising, since they do not require bulky set-ups and they are compatible with compact on-chip technology. The fascinating idea behind the use of these ``tuning knobs'' is to exploit the same semiconductor matrix that allows for the existence of QDs as the resource for solving problems related to their semiconducting nature itself. Progress in semiconductor technology has opened up the possibility to embed QDs in the intrinsic region of field-effect devices (such of n(p)-i-Schottky or n-i-p diodes) and to precisely control the electric field along the crystal growth direction (vertical electric field) by the simple application of a voltage \cite{bennett2010electric}. Similarly, the integration of semiconductor thin-films containing QDs onto piezoelectric substrates, such as lead-zirconate-titanate (PZT)~\cite{seidl2006effect} or lead magnesium niobate-lead titanate (PMN-PT)~\cite{seidl2006effect,zander2009epitaxial}, allows strain fields to be transferred to QDs via electrical means. Using the quantum-confined Stark-effect and variable deformation of the host semiconducting matrix, vertical electric fields and in-plane stress fields offer a precise and reversible way to engineer the QD electronic structure and have been instrumental in bringing into resonance the levels of QD molecules \cite{krenner2005direct,zallo2014strain}, to tune QD levels into resonance with cavity modes \cite{laucht2009dephasing,zander2009epitaxial,Lin2011stress}, and to control binding energies of excitonic complexes \cite{trotta2013independent,kaniber2011electrical,ding2010tuning}. Thanks to the broad-band tunability of the QD emission energy \cite{bennett2010electric,trotta2012nanomembrane} they have also enabled the first two experiments showing two-photon interference between remote QDs~\cite{patel2010two,Flagg2010interference}. Despite these impressive results and the tremendous efforts required to achieve them, having at hand single ``tuning knobs'' is very often not sufficient to meet some of the very stringent requirements set by advanced quantum optics experiments. A prominent example is represented by the difficulties encountered in erasing the FSS \cite{singh2010lower}. Theoretical and experimental results have demonstrated that the application of either stress or electric fields to single QDs generally results in a lower bound of the FSS (usually larger than $1~\mu$eV) caused by the coherent coupling of the two bright exciton states \cite{bennett2010electric,plumhof2011strain}. Only a few ``hero'' QDs can be tuned for entangled photon generation \cite{muller2009creating}, thereby hindering the implementation of QD photon sources in advanced quantum communication protocols \cite{Gisin2002Quantum}.

In this chapter, we describe a novel class of hybrid piezoelectric-semiconductor devices that allow large stress and electric fields to be \emph{simultaneously} applied to single semiconductor QDs. Despite the idea to combine independent fields emerges naturally from the need of a tighter control over the properties of the quantum emitters, it has been largely overlooked due to the common opinion that the use of several ``knobs'' simply extends the tunability of the QD emission properties. In strong contrast, we show that the effect of strain and electric field are complementary and that the resulting \emph{electro-elastic} field allows addressing tasks not solvable with  existing approaches. 

The chapter is divided as follows: In the first section we discuss the technological steps required to build up the hybrid semiconductor-piezoelectric devices capable of delivering electro-elastic fields to single QDs. We demonstrate immediately the technological relevance of this approach introducing a wavelength-tunable, high-speed and all-electrically-controlled source of single photons. In the following section we focus on the idea of using strain and electric field to achieve independent control of different QD parameters. In particular, we show independent control of (i) charge state and emission energy, (ii) exciton and biexciton energies, and (iii) amplitude and phase of mixing of bright exciton states. The latter achievement means that the electro-elastic fields can be used to tune any QD for the generation of highly polarization-entangled photon-pairs.

\begin{figure}[t]
\centering
\includegraphics[scale=0.85]{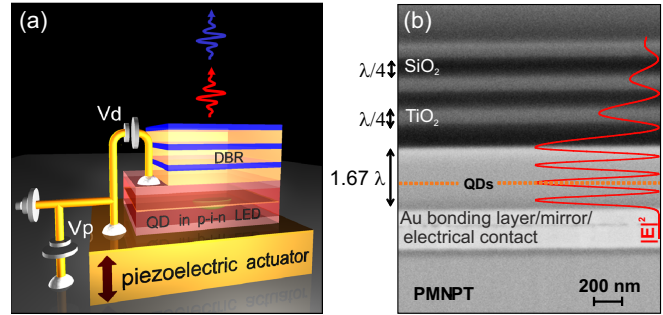} 
\caption{Strain-tunable quantum dot devices. (a). Sketch of the dual-knob device: A p-i-n nanomembrane containing self-assembled QDs is integrated on top of a piezoelectric actuator, allowing the {\em in situ} application of biaxial strain by tuning the voltage $V_p$. Electrons and holes are electrically injected by setting $V_d$. The top Distributed Bragg Reflector (DBR) completes a metal-semiconductor-dielectric planar cavity. (b). SEM image of a cross section of the device prepared by focused ion beam (FIB) cutting. The solid curve depicts the square modulus of the electric field inside the cavity. The QD layer is located one $\lambda$ below the DBR (see dashed line), in turn composed by three pairs of SiO$_{2}$/TiO$_{2}$.} 
\label{fig:Figure1}
\end{figure}

\section{Hybrid semiconductor-piezoelectric quantum dot devices: the first high-speed, wavelength-tunable, and all-electrically-controlled source of single photons}

The hybrid devices discussed here are obtained by merging the semiconductor and piezoelectric technologies. Diode-like (p-i-n or n-i-Schottky) nonanomembranes containing In(Ga)As QDs are integrated onto [Pb(Mg$_{1/3}$Nb$_{2/3}$)O$_{3}$]$_{0.72}$[PbTiO$_{3}$]$_{0.28}$ (PMN-PT) piezoelectric actuators featuring giant piezoelectric response. For details about device fabrication we refer the reader to the specific paper~\cite{trotta2012nanomembrane}.

Fig. \ref{fig:Figure1} shows a sketch of the final device, which features two electrical tuning knobs: a voltage applied to the nanomembranes ($V_{d}$) allows the electric field ($F_{d}$) across the QDs (up to $\sim 200$~kV/cm) to be controlled, as in standard field-effect devices. Simultaneously, the application of a voltage ($V_{p}$) to the PMN-PT results in an out-of-plane electric field $F_p$ that leads to tensile or compressive in-plane strains (up to $\pm 0.2 \%$ at cryogenic temperatures) in the QD layer. Unlike PZT, PMN-PT is capable of larger in-plane strains, which is crucial for broadband tunability. It is also worth noting that the Au layer between the PMN-PT and the nanomembrane plays a threefold role here: It acts as a stiff strain-transfer layer, as an electrical contact for both the nanomembrane and the PMN-PT, and it represents the bottom mirror of a metal-semiconductor-dielectric-planar planar cavity \cite{huffaker1995resonant} featuring extraction efficiencies as high as $15 \%$. Finally, depending on the particular design of the diode (n-i-Schottky or n-i-p diode), magnitude and sign of $V_d$, the electric field can be used to control the QD energy levels via the quantum-confined Stark-effect, to control the charge state of the QD or to inject carriers electrically. The latter operation mode is described in the following, where we report on the realization of the first high-speed, energy-tunable, and all electrically-controlled sources of single photons. 

\begin{figure}[t]
\centering
\includegraphics[scale=0.60]{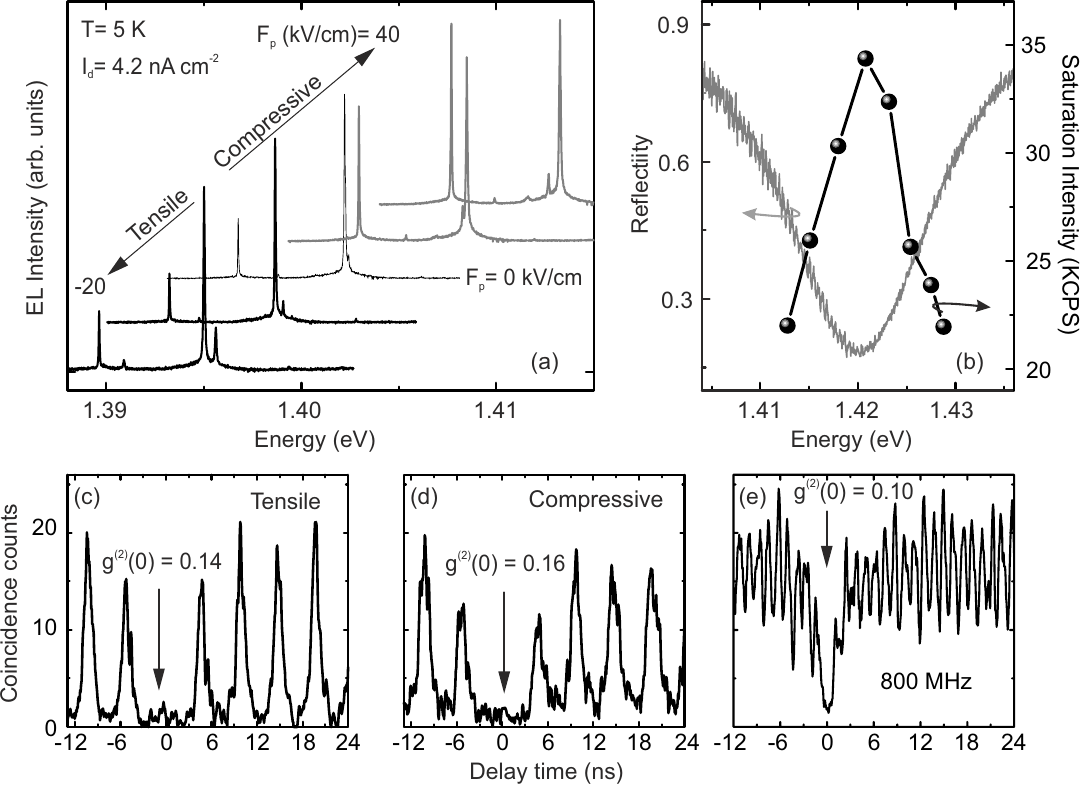}
\caption{\keyword{A wavelength-tunable, high-speed, bright, and all electrically controlled source of single photons.} (a). Low temperature ($T=5$~K) electroluminescence spectra of a single QD embedded in the dual-knob device of Fig. \ref{fig:Figure1} as a function of the electric field across the piezoelectric actuator and for a fixed current density $I_d$. Black thick (gray thin) lines correspond to tensile (compressive) strain. The spectrum obtained at $F_p=0$ is also reported (black line). (b). Reflectivity spectrum (gray line) for one of the devices. The points connected by a line represent the saturation intensity of the $X$ transition for different values of strain. (c). Autocorrelation measurements for an $X$ embedded in the device under tensile strain.  The QD is driven with $300$~ps long pulses at $200$~MHz. (d). Same as (c) for compressive strain. (e). Same as (c) for the unstrained case and at $800$~MHz.}
\label{fig:Figure2}
\end{figure}

In standard quantum-light-emitting diodes (Q-LEDs), InGaAs QDs are embedded in the intrinsic region of GaAs p-i-n device \cite{salter2010entangled}. When the applied bias exceeds the turn-on voltage, charge carriers are electrically injected into the QDs and photons of different frequencies are emitted in the recombination processes. In this operation mode, the electric field across the QDs is taken up to inject carriers and cannot be used in a trivial manner to modify sizably the QD electronic properties. Contrarily, the dual-knob device sketched in Fig. \ref{fig:Figure1}~\cite{trotta2012nanomembrane} addresses successfully this hurdle. Fig. \ref{fig:Figure2}a shows several \emph{electro-luminescence} spectra of a single QD as a function of the electric field across the piezoelectric actuator, i.e., as a function of in-plane biaxial strain. Spectra obtained for both tensile ($F_{p}$ down to $-20$~kV/cm) and compressive ($F_{p}$ up to $40$~kV/cm) strains are displayed to show that the QD emission lines ($X$, $XX$ and negative charge exciton, $X^{-}$) can be shifted in a $\sim 20$~meV-broad spectral range without loss of intensity or line broadening and, most importantly, during electrical injection. The reported tuning range is comparable with the inhomogeneous broadening of QD emission, meaning that any two QDs in the ensemble can be tuned into energetic resonance, an important prerequisite for transferring quantum information between independent quantum emitters via HOM-type interference \cite{hong1987measurement}. It is also interesting to note that during active deformation of the nanomembrane-LEDs, the frequency of the Fabry-Perot mode of the metal-semiconductor-dielectric planar cavity remains almost unaffected (shift up to 1~meV).  This ``pinning'' of the Fabry-Perot mode is interesting because one can design the cavity for a certain frequency and use strain to bring remote QDs into resonance in the spectral position where light extraction efficiency is maximized. In spite of the low quality factor of the cavity ($Q \approx 102$), Fig. \ref{fig:Figure2}c shows a clear enhancement of light extraction efficiency when the $X$ line is tuned trough the center of the cavity mode. Such a low quality factor can also be seen as an advantage for broad-band operation that allows for enhancing the different lines of the same QD (like $X$ and $XX$), which is fundamental for the efficient generation of photon pairs.

In order to prove that these novel devices deliver non-classical light, we performed autocorrelation measurements~\cite{hong1987measurement}. The coincidence counts recorded on the X transition of a single QD at $F_p= -10$~kV/cm, i.e., under applied tensile stress, are shown in Fig. \ref{fig:Figure2}c~\cite{zhang2013nanomembrane}. The periodic autocorrelation peaks\footnote{The electro-luminescence is excited by injecting short electrical pulses (pulse width less than $300$~ps, amplitude of $-0.7$~V and repetition rates up to $0.8$~GHz) superimposed to a direct current (DC) bias of $V_d = -1.7$~V, just below the turn-on voltage of the diode.} together with the absence of the peak at zero time-delay provide evidence of photon antibunching and of single-photon emission. The time separation between the neighboring peaks is $5$~ns, which corresponds to the repetition rate of $200$~MHz used for this experiment. While the normalized value of the second order correlation function at zero time delay $g^{(2)} (0)$ proves unambiguously that the source is a single quantum emitter ($g^{(2)} (0) <0.5$), it also shows a multi-photon emission probability of $0.12(2)$. Along with the background and the dark counts of the single photon detectors, this finite probability could originate from carrier recapture phenomena on a time scale comparable with the exciton lifetime, as observed in similar systems~\cite{bennett2006single}. Irrespective of the origin of the non-zero value of the $g^{(2)} (0) $, measurements performed at different $F_p$ (see Fig. \ref{fig:Figure2}d) show no significant change of the value of $g^{(2)} (0) $. This finding finally proves that the emission of single photons is not degraded by the application of such large stress fields to the LED and that our device can be used as an energy-tunable and bright source of single photons. In addition to the tunability in energy and the high extraction efficiency, our QD-LED allows for high-rate photon generation, another important requirement for high-data rate single photon applications \cite{Gisin2002Quantum}. Fig. \ref{fig:Figure2}e shows autocorrelation measurements of an exciton driven with a train of electrical pulses separated in time by $1.3$~ns (corresponding to a repetition rate of $0.8$~GHz). We observe a strong suppression of the peak at zero time-delay with $g^{(2)} (0)  = 0.12(2)$, similarly to what was found at lower repetition rates (see Fig. \ref{fig:Figure2}c-d). It is noticeable that the neighbouring peaks start to merge with each other, and that the use of higher repetition rates would result in higher values of the $g^{(2)} (0) $. In fact, the width of the peaks in the autocorrelation measurement is mainly determined by the total jitter on the time interval between the start and stop events registered by the correlation electronics. Considering the rise and the decay time of the exciton transition, a total time jitter of $1.37$~ns was estimated, which in turn leads to $2.74$~ns-broad peaks in the autocorrelation plot. Being $1.3$~ns ($2.6$~ns) the temporal distance between two (three) consecutive electrical pulses, the neighbouring peaks start to merge while the value of $g^{(2)} (0)$ remains almost unaffected. In order to further increase the speed of our single-photon source, different excitation schemes with appropriate DC bias should be used, so as to reduce the total time jitter via quantum tunnelling of charges out of the QD, an effect induced by band bending. On the other hand, the device concept can be adapted for the integration of high-Q cavities, where the Purcell effect can be instrumental for increasing not only the speed of the single photon source via reduced radiative recombination times but also its brightness \cite{nowak2014deterministic}.

\section{Independent control of different quantum  dot parameters via electro-elastic fields}

In the previous section we have shown that the electric field can be used to inject carriers electrically into QDs while strain is used to modify the energy of the emitted photons. Exciting possibilities, however, arise when electric field and strain are used in synergy to achieve independent control of different QD parameters, as described in the following experiments.

\subsection{Independent control of charge state and emission energy}

We now demonstrate \cite{Trottaunpubl} how it is possible to obtain independent control of charge state and emission energy in a single QD. In light of recent experiments demonstrating spin-photon entanglement\cite{gao2013quantum,gao2012observation}  (see chapter by McMahon and De Greve)using $X^-$ transitions in single QDs (excitonic complexes consisting of two electrons and one hole), this possibility is particularly relevant, since it could pave the way towards QD-based quantum networks where remote spins are entangled via HOM interference of the photons emitted by remote QDs during the radiative recombination of the $X^{-}$.

\begin{figure}[ht]
\centering
\includegraphics[scale=0.55]{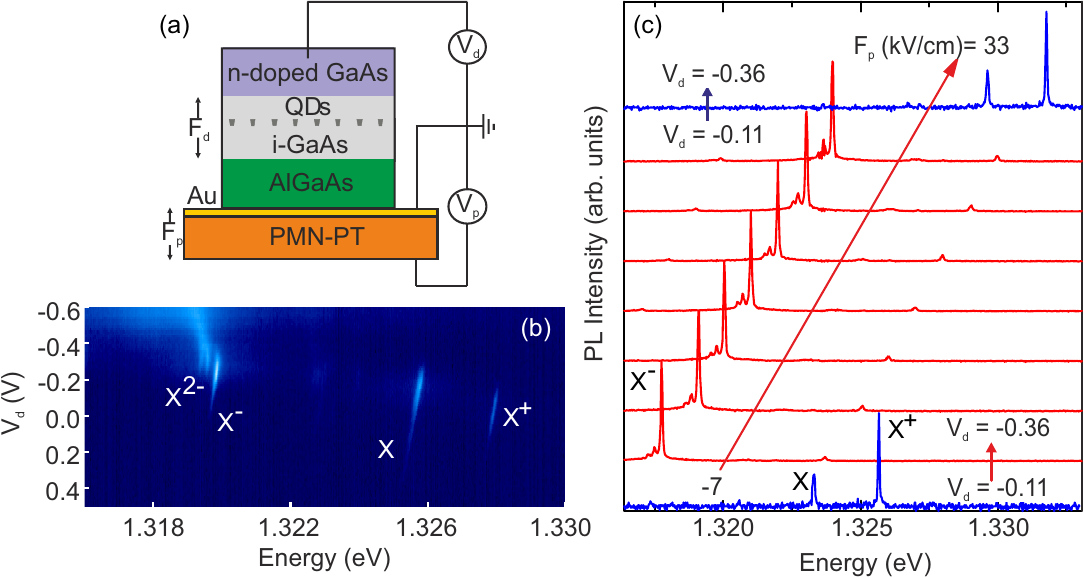}
\caption{\keyword{Independent control of charge state and emission energy in single QDs.} (a) Sketch of an n-i-Schottky diode containing QDs and integrated onto a PMN-PT actuator, similar as in Fig. \ref{fig:Figure1}(b). Micro-photoluminescence map (in false colour scale) of a single QD embedded in the device and as a function of $V_d$. Several recombination lines ascribable to exciton and charged exciton transitions can be clearly observed. (c). Micro-photoluminescence spectra of a single QD vs. $V_d$ and $V_p$. The latter values were chosen so as to demonstrate independent control of charge state and emission energy in a single QD. }
\label{fig:Figure3}
\end{figure}
For this experiment, we integrate n-i-Schottky diodes (instead of p-i-n LEDs) onto PMN-PT (see sketch of Fig.~\ref{fig:Figure3}a). As shown by Warburton and co-workers \cite{warburton2000optical}, this type of diodes allows single electrons to be injected into the QD with single-electron accuracy\footnote{This takes place when the voltage across the diode shifts the confined levels of QDs below the Fermi energy of the n-doped layer.}. In our device this is demonstrated in Fig. \ref{fig:Figure3}b, where micro-photoluminescence ($\mu$-PL) spectra of a single QD are reported as a function of $V_d$. At positive $V_d$ (large electric fields) the $\mu$-PL spectrum is composed by two sharp transitions related to the X and the positively charged excition ($X^+$).
When the electric field across the structure is reduced, such that the lowest state of the QD is aligned with the Fermi level of the n-doped contact, a new peak at lower energy appears and it gains intensity with $V_d$ at the expenses of the $X$ and $X^+$ transitions. This finding can be easily explained taking into account the \emph{tunneling of a single electron} from the top n-layer into the QD, and the new low-energy transition can be ascribed to the negatively charged exciton ($X^-$). Subsequent charging events (trapping of additional electrons into the QD) can be clearly seen as $V_d$ is further decreased, and two additional lines, most likely related to the $X^{2-}$ \cite{warburton2000optical}, appear in the $\mu$-PL spectrum (two electrons are injected into the QD). At even lower electric fields the QD is flooded with electrons and the PL spectrum evolves in a broad band. The $\mu$-PL map reported in Fig. \ref{fig:Figure3}b contains a plethora of additional information because the energetic-splittings between the different excitonic complexes are a direct result of the combined effect of Coulomb interaction and quantum confinement. In addition, the field-induced shift of the QD emission lines provides information about the QD permanent dipole moment and polarizability~\cite{bennett2010giant}. However, here we are mainly interested in discussing the possibility of obtaining simultaneous control of charge state and emission energy. Fig.~\ref{fig:Figure3}c shows how this is done: $V_d$ is first tuned so as to charge the QD with one electron ($X^-$), see the two bottom spectra of Fig. \ref{fig:Figure3}c. $F_p$ is then used to tune the $X^-$ emission energy while $V_d$ is kept fixed, as shown by the red arrows of Fig. \ref{fig:Figure3}c. Having the QD in a different strain configuration (different emission energy), we can now discharge it again by independent tuning of $V_d$ at fixed $F_p$ (see the two topmost spectra). Obviously, the experiment can be repeated charging QDs with an additional electron by simply adjusting $V_d$ at the value required to observe $X^{2-}$. 

\subsection{Independent control of exciton and biexciton energy}

So far, we have used the electric field across the diode for tasks that cannot be addressed with static strains, i.e., to inject carriers electrically and to control the charge state of a QD. We now explore a third and more interesting possibility: we use both fields to reshape the interaction energies among carriers confined in single a QD without affecting the energy of the QD fundamental excitation, i.e, the neutral exciton, $X$. In particular, we demonstrate for the first time independent and broad-range control of $X$ and $XX$ energy in a single QD. On the one hand, the demonstrated possibility to achieve color coincidence between $X$ and $XX$ photons could allow for testing the degree of entanglement of photon pairs produced using the recently proposed and not yet experimentally demonstrated time reordering-scheme~\cite{avron2008entanglement}. On the other hand, the broad range control over $X$ and $XX$ energies paves the way towards the development of energy-tunable sources of entangled photons via the time-bin scheme~\cite{jayakumar2014time}, and it would allow entanglement swapping experiments between distant QD-based qubits to be performed. 

\begin{figure}[ht]
\centering
\includegraphics[scale=0.65]{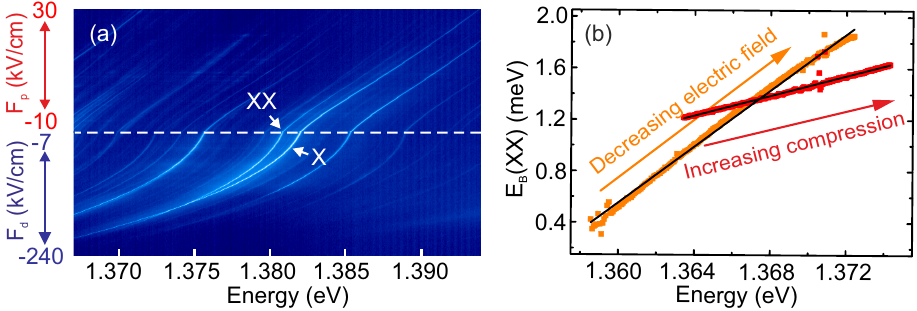}
\caption{\keyword{Electro-elastic control of excitons in QDs: ``additive mode'' operation.} (a) Micro-photoluminesnece map (in a false colour scale) of a single QD under the influence of large electric (bottom part) and strain (top part) fields. Recombination lines ascribed to the exciton ($X$) and biexciton ($XX$) transitions are indicated. (b). Biexciton binding energy $E_B(XX)$ of a single QD as a function of the energy of the exciton transition for strain (red points) and electric fields (orange points). The black lines are linear fits to the experimental data.}
\label{fig:Figure4}
\end{figure}

In the following experiments, we employ p-i-n diodes containing Al$_{0.4}$Ga$_{0.6}$As barriers surrounding a $10$~nm-thick GaAs quantum well that, in turn, hosts the QDs. The presence of the AlGaAs barriers reduces carrier ionization at very high electric fields \cite{bennett2010giant} and allows the QD emission lines to be shifted in a very broad spectral range when the diode is driven in reverse bias. When combined with the broad-band tunability provided by strain, this device offers unprecedented control over the QD emission properties, as described in the following \cite{trotta2013independent}. 
Fig. \ref{fig:Figure4}a shows the ``additive mode'' operation of the device, where strain and electric fields are used in sequence (first $F_d$ and then $F_p$) to shift the QD emission lines in the same direction.  From the $\mu$-PL map of Fig. \ref{fig:Figure4}a two important features of the device can be readily noticed: (i) The application of stress and electric fields results in a broad-range control of the QD emission lines, which can be as high as $40$~meV, one of the largest shift ever reported so far; (ii) The energy separation between the exciton ($X$) and biexciton ($XX$) emission lines -- which we refer to as the relative biexciton binding energy, defined as $E_B(XX)= E_X-E_{XX}$, where $E_{X,XX}$ indicate the emission energies -- is changing at different rates under strain and electric field. The different rates (almost a factor 2) can be better appreciated when $E_B (XX)$ is plotted against  $E_X$, as shown in Fig. \ref{fig:Figure4}b. While (i) is an expected result, (ii) is an interesting finding since it suggests that the two fields have a different effect on the interaction energies among carriers confined in the same QD. On the one hand, it has been shown \cite{ding2010tuning} that in-plane compressive biaxial strain (increasing exciton energy) increases the confinement potential of electrons and, consequently, their Coulomb repulsion while it leaves holes almost unaffected. On the other hand, an increasing vertical electric field (decreasing exciton energy) pulls electrons and holes apart and, since holes are more localized, the Coulomb repulsion increases much faster for them than for electrons.
For a detailed analysis of the field-induced changes of the Couloumb integrals, we refer the reader to the specific paper~ \cite{trotta2013independent}. Here, we would like to show that the different physical effects produced by the two fields can be used to reshape the electronic properties of single QDs so as to achieve independent control of the $X$ and $XX$ emission energy. To do so, we operate our dual knob device in ``subtractive mode'', i.e., strain and electric fields are used to shift the energy of the QD emission lines in opposite directions. 

\begin{figure}[ht]
\centering
\includegraphics[scale=0.65]{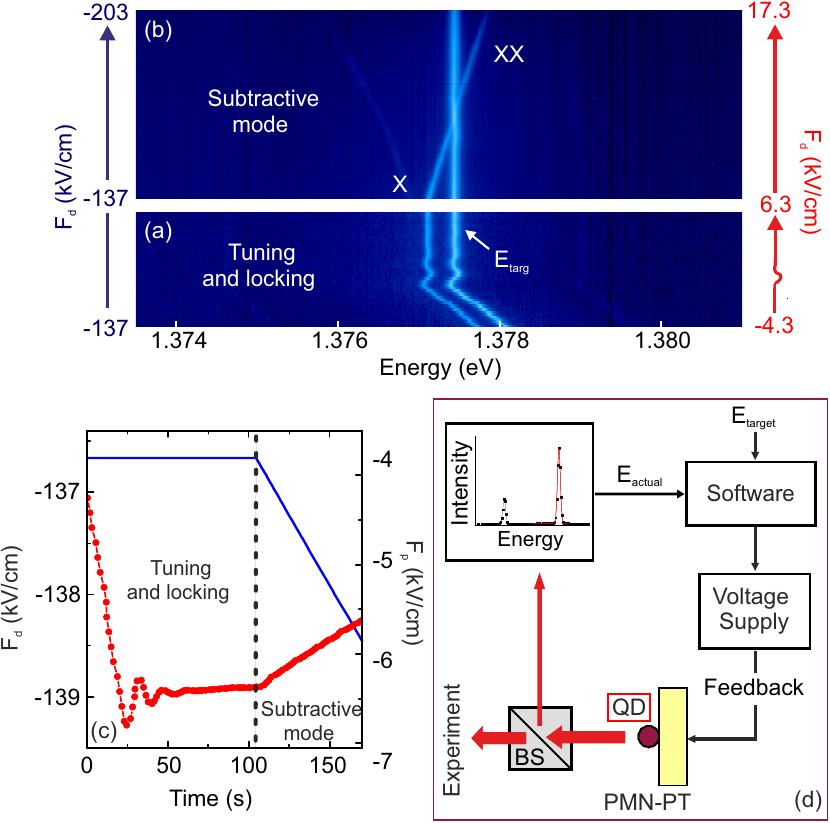}
\caption{\keyword{Electro-elastic control of excitons in QDs: ``Subtractive mode'' operation} (a). Color-coded $\mu$-PL map of a single QD whose $X$ is first tuned to the target energy of $E_{targ} = 1.3774$~eV and then locked at this value via $F_p$. During the experiment, $F_d \approx -137$~kV/cm. (b). Color-coded $\mu$-PL map of the same QD when the magnitude of $F_d$ is ramped up, while the exciton transition is locked at $E_{targ}$ via $F_p$, as explained in the main text. (c). Values of $F_d$ (blue) and $F_p$ (red) as a function of time as recorded during the experiment described in (a) and (b). (d). Sketch of the closed-loop system used to drive the emission energy of a QD to $E_{targ}$ and for energy stabilization via active feedback. The actual emission energy of the line to be stabilized is first obtained by fitting the $\mu$-PL spectrum and then compared to the $E_{targ}$ to provide feedback to the piezoelectric actuator. For details see Ref. \cite{trotta2012nanomembrane}.}
\label{fig:Figure5}
\end{figure}

Fig. \ref{fig:Figure5}a first shows the tuning and locking of the $X$ energy to the user-defined target energy ($E_{targ}$). This is achieved using a closed-loop feedback on the piezo-actuator that is capable to stabilize the $X$ frequency with $1~\mu$eV accuracy~\cite{trotta2012nanomembrane}, see Fig. \ref{fig:Figure5}d. After locking the $X$ to a recombination energy $E_{targ}$, we increase linearly the magnitude of $F_d$, while the exciton transition is kept fixed via $F_p$ [the time-evolution of both $F_p$ and $F_d$ is shown in Fig.~\ref{fig:Figure5}c]. In absence of the feedback, all the QD emission lines would redshift due to the quantum-confined Stark effect. However, as the $X$ shift is actively compensated by increasingly compressive strain and the two fields have a different effect on the $XX$ binding energy (see Fig. \ref{fig:Figure4}), we are able to change the spectral position of the $XX$ transition only. Remarkably, in the QD of Fig. \ref{fig:Figure4}b the $XX$ changes gradually from a binding to an antibinding configuration at fixed and predefined $X$ energy ($E_{targ}$). Thus, we are able for the first time \cite{trotta2013independent} to achieve independent control of the $XX$ and $X$ absolute energies. This confirms that the electro-elastic fields allow for reshaping of the interaction energies between carriers confined in a single QD without affecting the energy of the fundamental excitation in a QD, the neutral exciton. We believe that this result shows the real potential of our device, which is capable to address tasks inaccessible with single external fields used alone.

\section{Controlling and erasing the fine structure splitting for the generation of highly entangled photon pairs}

As mentioned in the introduction of this chapter, one of the unique features of our dual-knob device -- most probably the most important -- is its capability to erase the coherent coupling between the two bright excitonic states, that is, the exciton fine structure splitting (FSS). This is a fundamental requirement for the generation of polarization entangled photon pairs during the radiative decay of the biexciton ($XX$) to the exciton ($X$) to the crystal ground state, which has stimulated the efforts of researchers worldwide, who have struggled for more than 10 years after the first proposal \cite{PhysRevLett.84.2513} to find a reproducible way to suppress the FSS. In this section we address this issue in detail. In the first part~\cite{trotta2012universal} we discuss the theory underlying the FSS and we illustrate the reasons why our device is capable to correct for the structural asymmetries at the origin of the FSS. In the second part \cite{trotta2014highly} we demonstrate that at zero FSS QDs can generate photon pairs featuring a very high degree of polarization entanglement, high enough to violate Bell's inequality without the need of temporal and spectral filtering techniques.

\subsection{Controlling and erasing the exciton fine structure splitting via electro-elastic fields.}

The appearance of an energetic splitting between the two bright excitonic states is a manifestation of the spin-spin coupling of the electron and hole forming the exciton, i.e., the electron-hole exchange interaction~\cite{bayer2002fine}. Already at this point a simple question may arise: how can there be an exchange coupling between electrons and holes if this fundamental interaction involves indistinguishable particles? The answer is that in semiconductors physics what we call electron and hole are both perturbations of a many-electron system. In this sense an electron-hole exchange interaction exists, and the first clear report in semiconductors dates back to 1979, when W. Ekardt and co-workers reported on an accurate theoretical and experimental study of bulk GaAs and InP \cite{ekardt1979determination}. The determined splittings were found to be considerably small ($\approx 10~\mu$eV) and quite difficult to observe due to the broadening of the involved optical transitions. A few years later, R. Bauer et al. reported on the appearance of doublets in the PL spectra of heavy-hole excitons confined in GaAs/AlGaAs quantum wells \cite{Bauer1986Proced}, which they ascribed to the exchange coupling. Despite this study triggered a lively debate about the origin of these splittings \cite{blackwood1994exchange,Andreani1990Exchange}, two characteristic features of the exchange interaction became immediately clear: (i) The exchange interaction depends closely on the spatial extent of the exciton wavefunction, and is therefore expected to be enhanced in low dimensional systems; (ii) The symmetry of the excitons is what eventually determines the appearance of energetic splittings in the PL spectra and can be actually exploited to extract information about the microscopic structure of the system under study~\cite{blackwood1994exchange,van1988optically}. Considering (i), it is not surprising that the exchange interaction has a central role in QD physics, and its interplay with quantum confinement is vital for the understanding of the QD optical properties. The first experimental work showing exchange-induced splittings of ground state excitons in QDs appeared in 1996~\cite{gammon1996fine}, which also signs the date when the term fine structure splitting (FSS) appeared for the first time for QDs. Since then, many steps toward a complete understanding of the theory underlying the FSS were taken \cite{bayer2002fine,singh2010lower,seguin2005size}. Its existence suddenly became a ``problem'' in the early 2000, when C. Santori and co-workers \cite{santori2002polarization} showed that suppression of the FSS is fundamental for the efficient generation of polarization entangled photon pairs using the $XX$-$X$-$0$ radiative decay. As mentioned in the point (ii) above, the latter possibility is strongly connected to the symmetry of excitons confined in QD: Theoretical calculations show that the coherent coupling of the two bright excitonic states, and hence the FSS, appears every time the QD structural symmetry is lower than $D_{2d}$. Since this is the case even in ideal lens-shaped Stranski-Kranstanow QDs based on conventional zincblende semiconductors ($C_{2v}$ structural symmetry) alternative growth protocols capable to produce highly symmetric QDs have been developed \cite{juska2013towards,kuroda2013symmetric}. Despite impressive progress in this field, however, this approach fails to deliver a substantial number of QDs with FSS $< 1~\mu$eV. This fact can be qualitatively understood considering that there are inevitable fluctuations in the exact number of atoms composing QDs, their arrangement in the host matrix and intermixing with the substrate and the cap material~\cite{rastelli2008three}, thus rendering the possibility to grow semiconductor QDs showing specific properties of symmetry a mere theoretical construct. These difficulties have further stimulated the search for alternative routes relying on post-growth tuning of QD properties via the application of stress~\cite{seidl2006effect,kuklewicz2012electro}, electric~\cite{bennett2010electric}, and magnetic~\cite{stevenson2006semiconductor} fields. It turns out, however, that even with the aid of such ``tuning knobs'' it is extremely difficult to drive QD excitons towards a universal level crossing, mainly due to the coherent coupling of the two bright states~\cite{bennett2010electric,plumhof2011strain}. Suppression of the FSS can be instead achieved using two independent or at least not-equivalent external fields, as strain and electric field provided by our dual-knob device. Before showing how to achieve that experimentally, we discuss the relevant theory. Inspired by the work of M. Gong and co-workers~\cite{gong2011exciton}, we consider the combined effect of a vertical electric field ($F$) applied along the [001] crystal direction of GaAs and anisotropic biaxial stresses~\cite{kumar2011strain,kumar2014anomalous} of magnitude $p=p_1 - p_2$, where $p_1$ and $p_2$ are the magnitudes of two perpendicular stresses applied along arbitrary directions in the (001) plane. (Note that any in-plane stress configuration can be decomposed in such a way, with $p_{1,2}$ the principal stresses). The effective two-level Hamiltonian for the bright excitons takes the form~\cite{trotta2012universal}:

\begin{equation}
H_{ex} = \begin{pmatrix}
\eta + \alpha p + \beta F & k+ \gamma p \\
k+ \gamma p & -(\eta + \alpha p + \beta F) 
\end{pmatrix}
\label{eq:Trotta_eqMatrix}
\end{equation}

Two sets of parameters enter in $H_{ex}$: (i) the parameters $k$ and $\eta$, which are specific of every QD and account for the lowering of the structural symmetry down to $C_1$, i.e., the most generic case of QDs without structural symmetry; (ii) the parameters of the two perturbations $\alpha$, $\gamma$ and $\beta$. In particular, $\alpha$ and $\gamma$ account (via the elastic compliance constants renormalized by the valence band deformation potentials) for the direction of the applied stress while the parameter $\beta$ (proportional to half of the difference of the exciton dipole moments) is related to the electric field across the diode. Before proceeding further in the analysis it is important to point out that all these parameters combine together in two observables: the magnitude of the FSS and the polarization direction of the exciton emission $\theta$ \cite{gong2011exciton}. The latter parameter represents the orientation of the exciton eigenstates with respect to the directions of the underlying crystal.  Finally, diagonalization of the Hamiltonian in Eq. \ref{eq:Trotta_eqMatrix} gives the following values of $s$ and $\theta$: 

\begin{equation}
s= \left| \left( \eta + \alpha p + \beta F \right)^2  + \left( k + \gamma p \right)^2 \right| ;\  tan(\theta_{\pm}) = \frac{ k + \gamma p}{\eta + \alpha p + \beta F \pm s};
\label{eq:Trotta_fss}
\end{equation}

It can be easily shown that the expression for the FSS has always a minimum at zero when the magnitude of F and p take the values

\begin{equation}
p_{crit} = - \frac{k}{\gamma} \ \  \text{and} \ \  F_{crit} = \frac{\alpha k}{\gamma \beta}- \frac{\eta}{\beta}
\label{eq:Trotta_pcrit}
\end{equation}

In other words, there are \emph{always} values of strain and electric field such that $s=0$, regardless of the QD structural symmetry, i.e., regardless of the exact values of $\eta$ and $k$. At this point, it is important to discuss why exactly two external fields are needed to cancel the FSS and a lower bound for the FSS is instead systematically observed in experiments performed with single tuning knobs \cite{bennett2010electric,bennett2010giant}. Assuming that only one field acts on QDs, e.g. strain, one can minimize the expression of the FSS reported in equation \ref{eq:Trotta_fss} to find the value of $p$ leading to $s=0$, that is,$\frac{k}{\eta} = - \frac{2 \gamma}{\alpha}$. This equation obviously connects the parameters of the QDs ($\eta$ and $k$) to the parameter characterizing the external perturbation ($\gamma$ and $\alpha$). Since $\eta$ and $k$ are unknown and fluctuate from dot to dot, this equation also implies that the $s=0$ condition can be achieved only if one has \emph{active} control over the direction and magnitude of the applied stress field ($\gamma$ and $\alpha$). If this direction is instead fixed -- as in all the experiments performed so far -- a lower bound of the FSS is generally observed and only the QDs that happen to be ``just right'' for the chosen perturbation can be tuned to low FSS values. \emph{In strong contrast, equation \ref{eq:Trotta_fss} shows that the capability of two independent fields to erase the FSS does not depend on the details of the QD under study}. It is obvious that in real experiments large enough tuning ranges are needed to access the values of $p_{crit}$ and $F_{crit}$ given above. This requirement appears to be satisfied by our device, which allows us to tune systematically all the QDs we measure to $s=0$. 

\begin{figure}[htb]
\centering
\includegraphics[scale=0.65]{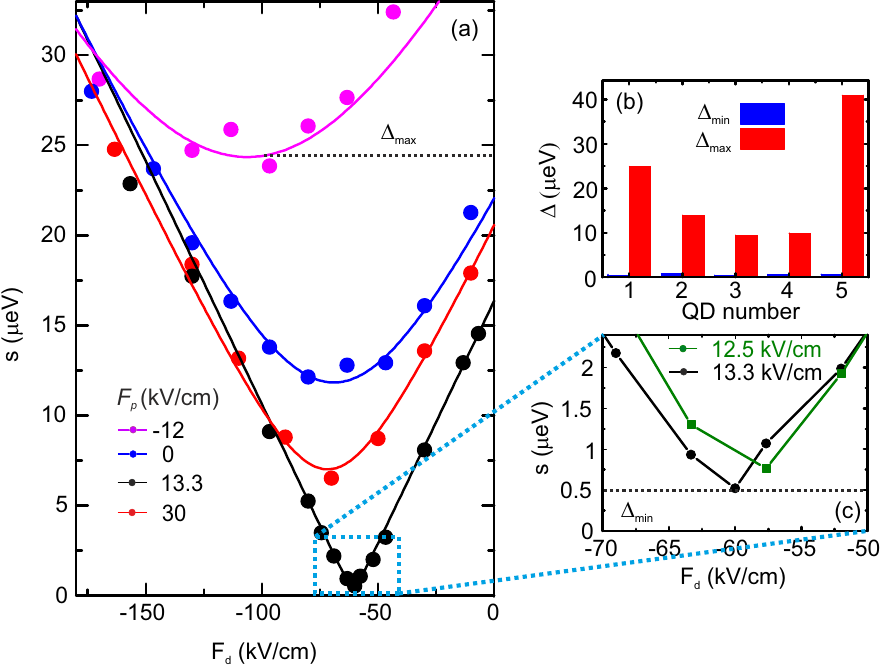}
\caption{\keyword{Universal recovery of the bright exciton level degeneracy} (a). Behavior of the FSS of a single QD as a function of $F_d$ and for different values of $F_p$. The solid lines are fits to the experimental data obtained using Eqs. \ref{eq:Trotta_fss},\ref{eq:Trotta_pcrit} with fixed QD parameters. $\Delta_{max}$ indicates the maximum value of $\Delta$ (lower bound for the FSS vs $F_d$) observed for this QD. (b) Zoom of (a) in the region of small FSS. The data taken at $F_p = 12.5$~kV/cm, not displayed in Fig. \ref{fig:Figure3}a, are also reported (green points connected by lines). $\Delta_{min}$ indicates the minimum value of $\Delta$ observed for this QD. (c). Histogram of $\Delta_{max}$ and $\Delta_{min}$ for 5 random QDs measured in the same device. }
\label{fig:Figure6}
\end{figure}

\begin{figure}[htb]
\centering
\includegraphics[scale=0.85]{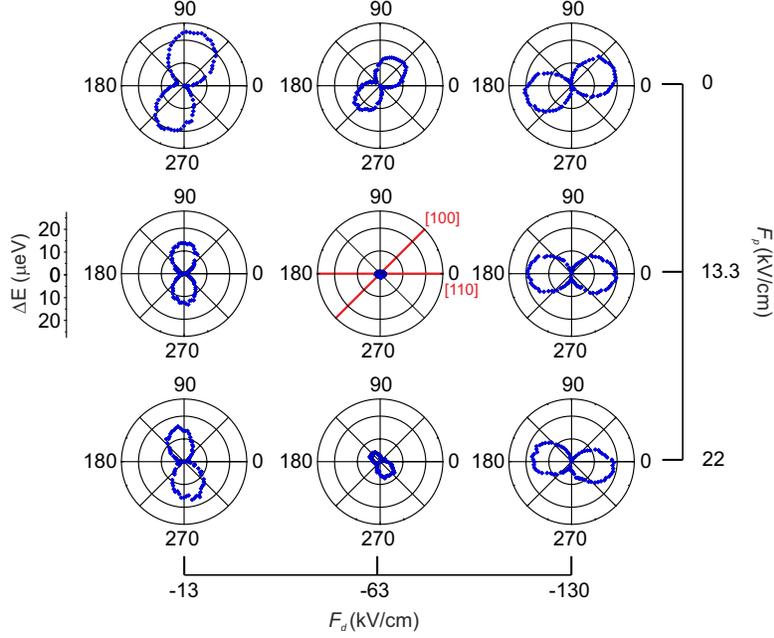}
\caption{\keyword{Independent control of exciton fine structure splitting and polarization angle} (a). Dependence (in polar coordinates) of $\Delta E$ vs. $\theta$, see text. The length and the orientation of the ``petals'' give the value of the FSS and $\theta$, respectively. Specific values of $F_p$ and $F_d$ around $FSS \approx 0$ have been used to construct this plot. In the central panel, the [110] and [100] crystal directions of the GaAs matrix hosting the QDs are also indicated.}
\label{fig:Figure7}
\end{figure}

Fig. \ref{fig:Figure6}a shows the behaviour of the FSS for a QD as a function of $F_d$ and for different values of $F_p$. In the tensile regime ($F_p < 0$), a lower bound ($\Delta$) for the FSS is observed. Under compressive strain ($F_p > 0$) $\Delta$ first decreases to $\sim 0.5~\mu$eV (see Fig. \ref{fig:Figure6}c), a value comparable with the experimental spectral resolution, and then increases again. Remarkably, the behaviour of the FSS against $F_d$ and $F_p$ is the same in all the measured QDs (see Fig. \ref{fig:Figure6}b) but for the specific values of $F_d$ and $F_p$ at which the FSS reaches $s=0$. We have used equation \ref{eq:Trotta_pcrit} to fit the experimental data (see solid lines in Fig. \ref{fig:Figure6}a) and we have found an excellent agreement. This confirms not only the existence of a universal method to tune the bright exciton states towards level crossing, but it also points out that the simple theory discussed above -- which neglects higher order terms in $p$ and $F$ -- is able to grasp the main features of the experiments. Additional information can be inferred looking at the behaviour of the polarization direction of the exciton emission ($\theta$) in relation with the FSS. Fig \ref{fig:Figure7} shows the dependence (in polar coordinates) of $\Delta E$ as a function of the angle the linear polarization analyser forms with the [110] crystal axis, where $\Delta E$ is half of the difference between $XX$ and $X$ energies minus its minimum value (see ref \cite{trotta2012universal} for details). In other words, the length and orientation of the petals give the value of $s$ and $\theta$, respectively. It is clear that when the eigenstates are oriented along the [110] (close to the [100]) or the perpendicular direction, the application of $F_d$ ($F_p$) leads to $s=0$. Since the electric field acts as an effective deformations along the [110] direction and the principal stress axis in this device is close to the [100] direction~\cite{trotta2012universal}, this implies that the excitonic degeneracy can be restored if one external perturbation (e.g. $F_d$) is used to align the polarization axis of the exciton emission along the axes of the second perturbation (e.g. $F_p$), which is then able to compensate \emph{completely} for the difference of the confining potentials of the two bright exciton eigenstates, i.e., is able to tune the FSS to zero. This finally shows that the possibility to suppress the FSS is intrinsically connected to the capability of our dual-knob device to achieve independent control of the magnitude of the FSS and the polarization direction of the exciton emission, $\theta$.

\subsection{Generation of highly entangled photon pairs via electro-elastic tuning of single semiconductor QDs. }

\begin{figure}[htb]
\centering
\includegraphics[scale=1]{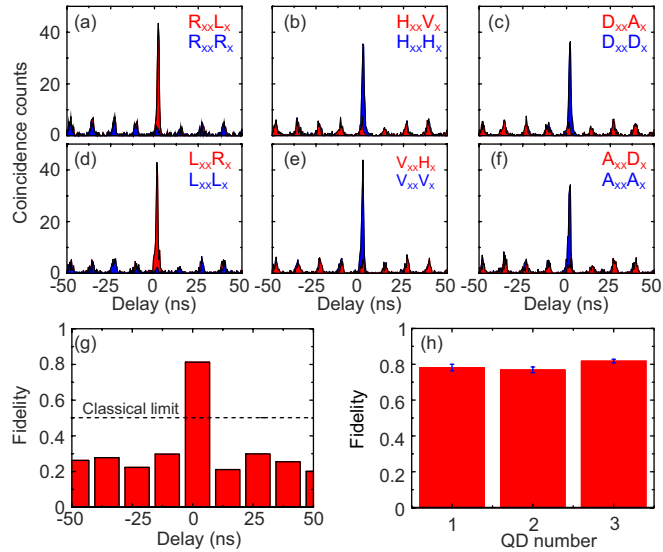}
\caption{\keyword{Characterization of the entanglement.} (a)-(f).  Projective measurements in the polarization basis. The peak at zero delay shows coincidence counts between exciton and biexciton photons for a QD whose FSS has been tuned via the electro-elastic field to zero. The left, central, and right panels correspond respectively to the circular, linear, and diagonal basis. (g). Fidelity vs. time delay as obtained from the analysis of the data shown in in panels (a)-(f). The dashed line indicates the classical limit. (h). Fidelity values (at zero time delay) for three different QDs from the same device.}
\label{fig:Figure8}
\end{figure}

\begin{figure}[ht]
\centering
\includegraphics[scale=0.85]{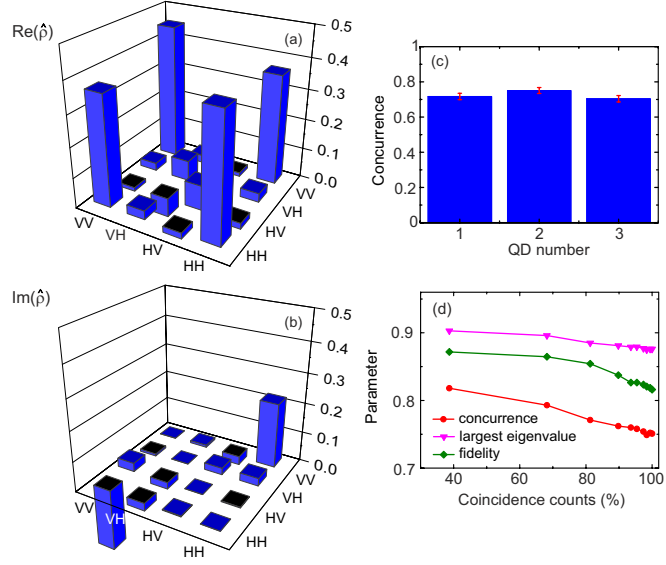}
\caption{\keyword{Quantification of the entanglement degree and analysis with temporal post-selection.} (a) Real and (b) imaginary part of the two-photon state density matrix as obtained in measurements on a single QD via quantum state tomography. (c). Concurrence values for three different QDs. (d). Temporal post-selection of the parameters characterizing entanglement (concurrence, fidelity, and largest eigenvalue) as a function of the fraction of the recorded coincidence counts (normalized to the total counts calculated for $w=10$~ns). }
\label{fig:Figure9}
\end{figure}

In the previous section we have demonstrated a universal method to systematically correct for the structural asymmetries that cause the FSS. This opens up the possibility to use arbitrary QDs for the generation of highly entangled photon pairs. When the intermediate $X$ states are degenerate ($s=0$), photon pairs produced during the $XX$-$X$-$0$ radiative decay are predicted to be in the maximally entangled Bell state $\psi = \frac{1}{\sqrt{2}} \left( H_{XX}H_X + V_{XX}V_X \right) $. Fig. \ref{fig:Figure8} shows $XX$-$X$ cross-correlation measurements for circular (Fig.~\ref{fig:Figure8}a, d), linear (Fig. \ref{fig:Figure8}b, e) and diagonal (Fig. \ref{fig:Figure8}c, f) polarization basis for a QD whose FSS has been tuned to $s = (0.2 \pm 0.3)~\mu$eV. We observe strong correlation when recording coincidence events for the following projections: $H_{XX} H_X$ (or $V_{XX} V_X$), $D_{XX} D_X$ (or $A_{XX} A_X$) and $R_{XX} L_X$ (or $L_{XX} R_X$). On the other hand, the correlation peaks disappear for $H_{XX} V_X$ (or $V_{XX} H_X$), $D_{XX} A_X$ (or $A_{XX} D_X$) and $R_{XX} R_X$ (or $L_{XX} L_X$).
This is exactly the predicted behaviour of photon pairs emitted in the maximally entangled Bell state $\psi$. By integrating over all the events in the correlation peak at zero time delay it is possible to calculate the correlation visibilities $C_{AB}$, defined as the difference between co-polarized and cross-polarized correlations divided by their sum. We find $C_{HV}= 0.72(5), \left| C_{RL} \right| = 0.82(2), C_{DA} = 0.72(5)$. From the correlation visibilities, we can in turn calculate the fidelity to $\psi$ via the following formula \cite{hudson2007coherence} $f= \left( 1 + C_{HV} + C_{DA} + \left| C_{RL} \right| \right) / 4$  and obtain $f = 0.82(4)$ (see Fig. \ref{fig:Figure8}g). We have repeated the same measurements in three different QDs tuned to $s=0$ and we have found very similar values for the fidelity (see Fig. \ref{fig:Figure8}h). Since these values are always much larger than the classical limit of $f=0.5$, the experimental data clearly indicate that our strain-tunable device is capable of delivering polarization-entangled photon pairs. However, the fidelity to the Bell state is only an indicator of entanglement and cannot be used to obtain a quantitative estimate. For this reason, we have performed state tomography \cite{james2001measurement} and reconstructed the density matrix, $\hat{\rho}$ , of the two-photon entangled state. The real and imaginary parts of $\hat{\rho}$ for a selected QD are displayed in Fig. \ref{fig:Figure9}a-b. The matrix clearly satisfies the Peres inseparability criterion \cite{peres1996separability} for entanglement, being $-0.35 (<0)$ the minimum eigenvalue of its partial transpose. The $\hat{\rho}$ contains also imaginary components (see Fig. \ref{fig:Figure9}b), which point out to the presence of a phase delay between $\left| H_{XX} H_X \right\rangle$ and $\left| V_{XX} V_X \right\rangle$. Therefore, the state is not exactly the maximally entangled Bell state $\psi$ but rather $\psi^* \approx \frac{1}{\sqrt{2}} \left( \left| H_{XX} H_X \right\rangle + e^{-i(0.23 \pi)} \left| V_{XX} V_X \right\rangle \right)$ , which corresponds to the largest eigenvalue $\lambda = 0.86$ of $\hat{\rho}$ . In order to quantify the level of entanglement we have extracted from $\hat{\rho}$ the following metrics \cite{james2001measurement}: tangle ($T$), concurrence ($C$) and entanglement of formation ($E_F$). For the best QD studied here we obtain $T=0.56(3), C=0.75(2), E_F =0.66(5)$, but very similar values were found for other QDs (see Fig. \ref{fig:Figure9}c). 

It is important to point out that raw data were used in the analysis, without any background subtraction. The measured values are very high compared to previous results using QD-based photon sources \cite{PhysRevLett.96.130501,stevenson2006semiconductor,dousse2010ultrabright,juska2013towards,kuroda2013symmetric,Hafenbrak2007Triggered} and, most importantly, for an electrically-controlled optoelectronic device \cite{salter2010entangled,bennett2010electric,ghali2012generation}. In spite of this achievement, the level of entanglement is not yet ideal and mainly limited by depolarization of the exciton state. This, in turn, can be ascribed to two main mechanisms: (i) fluctuating QD environment \cite{stevenson2008evolution} and (ii) recapture processes \cite{dousse2010ultrabright}. Point (i) is related to local variations of magnetic and electric fields experienced by the QDs and produced, respectively, by the  nuclei of the QD material and by random charges. These fluctuating fields induce a variation of the FSS over time scales much faster than the time required to perform a state tomography, which can therefore reveal $s \approx 0$ only on average. Point (ii) is instead associated with processes in which the intermediate X level is re-excited to the XX level before it decays to the ground state. This mechanism can be optically driven or due to charged carriers trapped in the QD surrounding and produce background photons lowering the correlation visibilities. Temporal post-selection of the emitted photons \cite{salter2010entangled,dousse2010ultrabright,young2009bell} can be used to alleviate the deleterious effects just discussed, thought at expenses of the brightness of the entangled photon source. We have investigated this strategy in our QDs gradually reducing the temporal window $w$ we choose to integrate the correlation counts later used in the analysis. More specifically, we symmetrically discard photon-pairs arriving at longer positive and negative time delays. This is reasonable because recapture processes produce uncorrelated photons at negative time delays, while in the presence of a fluctuating FSS photons arriving at longer time delays are expected to exhibit lower fidelity to the Bell state. Fig. \ref{fig:Figure9}d shows the evolution of the concurrence ($C$), fidelity ($f$) and largest eigenvalue ($\lambda$) as a function of the fraction of coincidence counts (normalized to the total counts) recorded during the analysis. The different points correspond to different $w$, ranging from $10$ to $1$~ns, being the latter value close to the temporal resolution of the experimental set-up ($\sim 500$~ps). A monotonic increase of all the parameters quantifying entanglement can be clearly observed: we first note a slight increase of the parameters for $4$~ns~$<~w<10$~ns when less than $10\%$ coincidence counts are discarded. This behaviour can be easily explained considering the temporal width of the coincidence peak ($\sim 4$~ns), and points out the small, albeit deleterious, effect of background photons. A more pronounced effect is instead observed for $w<4$~ns: When $\sim 60\%$ of the counts are discarded ($w=1$~ns) a concurrence as high as $0.82$ is measured. We believe that the concurrence of our source can be further improved using resonant excitation techniques \cite{muller2014demand,jayakumar2013deterministic} and faster photon detectors.

The level of entanglement already achieved is particularly significant because it allows us to overcome the Bell limit \emph{without} post-selection of the emitted photons. A first indication of such a possibility is indicated by the level of concurrence. As a rule of thumb, a concurrence of $C\approx 0.7$ (or equivalently a tangle of $T \approx 0.5$) is necessary to violate Bell's inequality \cite{young2009bell}.

\begin{figure}[ht]
\centering
\includegraphics[scale=0.65]{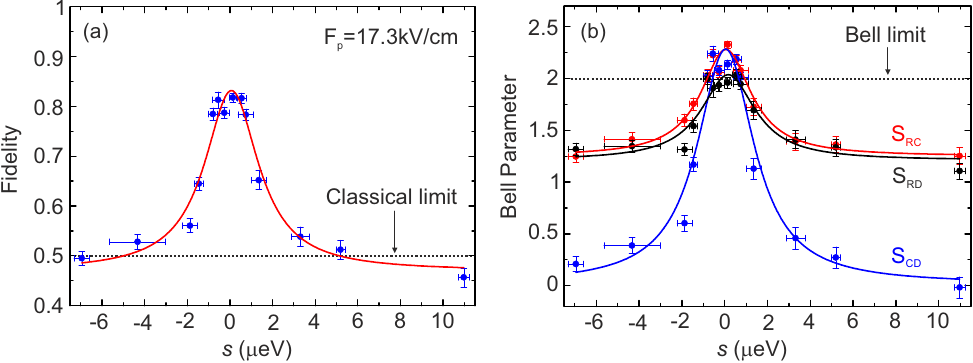}
\caption{\keyword{Classical vs. Bell limit.} (a). Fidelity values for a single QD while the FSS is driven through zero via the electro-elastic fields. The classical limit is indicated by the dashed line. The red line is a fit to the experimental data using a Lorentzian function. (b). Same as in (a) for the three different Bell's parameters introduced in the main text. The Bell limit is indicated by the dashed line.  }
\label{fig:Figure10}
\end{figure}

The values we measured in all our QDs are above this threshold (see Fig. \ref{fig:Figure9}c). Considering the unique capability of our device to drive the FSS through zero, it is extremely interesting to investigate for which values of the FSS we are actually able to violate Bell's inequality. Following references \cite{salter2010entangled} and \cite{young2009bell} we use the following equations for three different planes of the Poincar\'e sphere: $S_{RC} = \sqrt{2} \left( C_{HV} - C_{RL}\right),\  S_{RD} = \sqrt{2} \left( C_{HV} - C_{DA}\right),\  S_{CD} = \sqrt{2} \left( C_{DA} - C_{RL}\right)$. A value larger than 2 of one of these parameters ensures violation the Bell's inequality. Fig. \ref{fig:Figure10} shows the evolution of these parameters as well as of the fidelity as a function of $s$. The fidelity (see Fig.~\ref{fig:Figure10}a) first increases, it reaches the non-classical value of $0.82(4)$ and then it decreases again. As shown by Hudson and co-workers \cite{hudson2007coherence}, this behaviour can be approximated by a Lorentzian function, whose full-width at half maximum ($\sim 3~\mu$eV) nicely matches the lifetime of the exciton transition ($\tau \approx 1$~ns, being roughly constant for the range of electric fields explored during the experiment). This proves that significant entanglement can be measured once the FSS is reduced below the radiative linewidth of the $X$ transition. Violation of the Bell's inequality in all three planes (see Fig. \ref{fig:Figure10}b), however, can be achieved only for a very small range of FSS, $s<1~\mu$eV, where we measure $S_{RD} = 2.04(0.05),\  S_{CD} = 2.24(0.07),\  S_{RC} = 2.33(0.04)$. The latter parameter shows violation of Bell's inequality by more than 8 standard deviations and proves unambiguously that our electrically-driven source can produce non-local states of light. The values of the three Bell parameters further increase well above the limit of 2 using temporal post-selection of the emitted photons, with maximum values of $S_{RD} = 2.22(0.05),\  S_{CD} = 2.50(0.07),\  S_{RC} = 2.43(0.04)$ see ref. \cite{trotta2014highly}.

To conclude, we further stress the relevance of our results in the perspective of using QDs entanglement resources for applications: It is commonly believed that entanglement is quite tolerant to the presence of a small FSS. Fig. \ref{fig:Figure3}b readily confirms this general idea by showing that the classical limit can be beaten already for FSS$\approx 3~\mu$eV. However, overcoming the classical limit is not sufficient for applications relying on non-local correlations between the emitted photons, such as quantum cryptography and entanglement swapping \cite{Gisin2002Quantum,pan1998experimental}. One possible criterion to define the ``useful'' entanglement degree is the violation of Bell's inequality, initially proposed to demonstrate the entanglement non-locality and then used as a base for quantum cryptography \cite{ekert1991quantum}. Figure \ref{fig:Figure10}b clearly shows that this can be achieved only for FSS$< 1~\mu$eV, thus ultimately proving the importance of having at hand broad-band ``tuning knobs'' capable to suppress the FSS \cite{trotta2012universal}.

\section{Conclusions and outlook}

In this Chapter, we have introduced a novel class of semiconductor-piezoelectric devices that allows -- via the simultaneous application of strain and electric field -- for unprecedented control over the electronic and optical properties of self-assembled semiconductor quantum dots. The motivation behind the development of these devices is to use the same semiconductor matrix which allows for the existence of QDs as resource for solving some of the problems arising from their semiconducting nature itself, and in particular those that are hampering their exploitation as sources of non-classical light. In doing so, we have demonstrated the first all-electrically controlled (LED), wavelength-tunable (up to $20$~meV), frequency stabilized (down to $1~\mu$eV), high-speed (up to $0.8$~GHz) source of single photons. Most importantly, we have shown that the electro-elastic fields generated by our device are able to correct for the structural asymmetries that usually prevent QDs from emitting high-quality polarization entangled photon pairs. We believe that this dual-knob device opens up new frontiers for using QDs in quantum communication science and technology. In particular, it could be exploited in applications based on entanglement non-locality, such as quantum cryptography, as it allows using \emph{any arbitrary} QD to generate triggered entangled photon-pairs featuring high entanglement degree -- high enough to violate Bell's inequality without the need of inefficient temporal and spectral filtering techniques. Furthermore, our key idea of combining different external perturbations to achieve independent control of different QD parameters can be further extended in protocols focusing on the distribution of entanglement over the distant nodes of a quantum network, as in quantum relays and repeaters. In these applications, \emph{energy-tunable sources of entangled photons} are needed to match the color of the entangled photons emitted by remote QDs. This is a crucial prerequisite for teleporting entanglement via Hong-Ou-Mandel type two-photon interference. At present, this is out of reach even by the dual-knob device we have discussed in this chapter. In fact, the constraint of exciton level degeneracy requires specific values of strain and electric fields and, as a consequence, specific energies of the X or XX transitions. A different device concept is therefore needed to realize an energy-tunable source of entangled photons. However, having recognized that each QD parameter to control ``requires'' an independent external field, it is not difficult to envisage that external fields featuring three (or more) independent degrees of freedom will be the key to successfully address this task. We leave this point to future studies \cite{trotta2014energy}.

\section*{Acknowledgements} %
\begin{acknowledgement}
We thank O.~G. Schmidt who continuously supported the different stages of the research presented here, partially performed under his direction at the Institute of Integrative Nanoscience, IFW Dresden. We gratefully acknowledge P.~Atkinson and E.~Zallo for sample design and growth, and C. Ortix for theoretical support. We also thank J.~Wildmann, J.~X.~Zhang., J.~D. Plumhof, S.~Kumar, F.~Ding, J.~Schr\"oter, R.~O.~Rezaev, and E.~Magerl for their important contribution to the work and B.~Eichler, R.~Engelhardt, F.~Binder, S. Br\"auer, A.~Halilovic, E.~Vorhauer, U.~Kainz for invaluable support and technical assistance and K. D\"orr and A. Herklotz for help with the piezoelectric actuators. The work was supported financially by the European Union Seventh Framework Programme 209 (FP7/2007-2013) under Grant Agreement No. 601126 210 (HANAS), and by Bundesministerium f\"ur Bildung und Forschung (BMBF) project QuaHL-Rep (Contract No. 01BQ1032).
\end{acknowledgement}

\bibliographystyle{SpringerPhysMWM} 
\bibliography{TrottaBib}

\printindex
\end{document}